%% file: main.tex
\documentclass{article}

\usepackage[switch]{lineno}
\input{preamble_2.tex}
\usepackage{amsmath}
\usepackage{amsfonts}
\usepackage{spconf,amsmath,graphicx}

\usepackage[T1]{fontenc}
\usepackage{tempora} 
\ninept

\usepackage{amsmath,amssymb,amsfonts}
\usepackage{algorithmic}
\usepackage{algorithm2e}
\RestyleAlgo{ruled}
\usepackage{array}
\usepackage{textcomp}
\usepackage{stfloats}
\usepackage{url}
\usepackage{verbatim}
\usepackage{graphicx}
\usepackage{cite}
\usepackage{subcaption}
\usepackage{multirow}
\usepackage{xcolor}

\usepackage[scaled=.80]{beramono}

\usepackage{amsmath,amsthm,amssymb}

\input{AB_Preamble.tex}

\title{\vspace{-2.5cm}Unlimited Sampling Radar: a Real-Time End-to-End Demonstrator}
\name{Thomas Feuillen$^{1,2}$, Bhavani Shankar MRR$^{1}$ and Ayush Bhandari$^{2}$
\thanks{\scriptsize{The work of TF and AB is supported by the UK Research and Innovation council's FLF Program ``Sensing Beyond Barriers'' (MRC Fellowship award no.~MR/S034897/1). Further details on {Unlimited Sensing} and upcoming materials on \textit{reproducible research} are available via  \href{https://bit.ly/USF-Link}{\texttt{https://bit.ly/USF-Link}}. The work of TF and BSMRR was supported in part by the Luxembourg National Research Fund (FNR) through the CORE project SPRINGER under grant  C18/IS/12734677/SPRINGER.}}}
\address{$^{1}$ SPARC group, SnT, University of Luxembourg\\
$^{2}$  Department of Electrical and Electronic Engineering, Imperial College London}
\begin{document}
%
\maketitle
\begin{abstract}
In this paper, the trade-off between the quantization noise and the dynamic range of ADCs used to acquire radar signals is revisited using the Unlimited Sensing Framework (USF) in a  practical setting. Trade-offs between saturation and resolution arise in many applications, like radar, where sensors acquire signals which exhibit a high degree of variability in amplitude. 
To solve this issue, we propose the use of the co-design approach of the USF which acquires folded version of the signal of interest and leverages its structure to reconstruct it after its acquisition.
We demonstrate that this method outperforms other standard acquisition methods for Doppler radars.
Taking our theory all the way to practice, we develop a prototype USF-enabled Doppler Radar and show the clear benefits of our method. In each experiment, we show that using the USF increases sensitivity compared to a classic acquisition approach.
\end{abstract}
\vspace{-0.2cm}
\section{Introduction}
\vspace{-0.25cm}
Digital signal acquisition is the driving force behind all modern-day electronic technology. Digital acquisition is made possible via so-called \textit{Analog to Digital Converters} (ADCs).
Due to the nature of such devices, there is a fundamental trade-off between the dynamic range and the number of bits which is inevitable.
This problem is typically pervasive in ranging/localization applications because the dynamic range of the measurements is correlated with their spatial location.
Radar sensors, because of the radar equation that represents how the power of targets behave, are especially susceptible to this effect. 
Indeed, similar targets can be perceived by the radar with widely different amplitude, resulting in a need for high fidelity acquisition that can accommodate both \textit{strong} and \textit{weak} targets simultaneously.

\bpara{Related Work on Near-Far Problem.} Several strategies for the acquisition chain have been proposed to partially alleviate these limitations, from dithering before quantization \cite{xu2018quantized,rapp, Stockham }, to other types of ADC architecture \cite{9264222, 7338380}.
That said, the gain they provide entails higher costs both in terms of implementation and  sampling rate (thus affecting power consumption and bit budget), thus making them unappealing solutions.
We attribute such drawbacks to the fact that in most of such approaches, hardware and algorithms are detached from each other.

\bpara{From Computational Sensing to Computational Radars.} To overcome the dynamic range bottleneck in the conventional digital acquisition pipeline, the \textit{Unlimited Sensing Framework} (USF) \cite{Bhandari:2017:C,Bhandari:2020:Ja,Bhandari:2020a,Bhandari:2021:J} has been recently proposed in the literature.
The USF is based on a co-design of hardware and algorithms.
The continuous-time input signal is folded via modulo non-linearity before sampling.
This avoids any risk of ADC saturation.
Advanced signal processing methods then algorithmically \textit{unfold} the modulo samples, thus enabling high-dynamic-range signal recovery.
Experiments based on the modulo ADC hardware \cite{Bhandari:2021:J} have shown that signals as large as $25\times$ to $30\times$ the ADC's maximum voltage can be recovered in practice. 
\begin{figure}[!t]
    \centering
\includegraphics[width=0.45\textwidth]{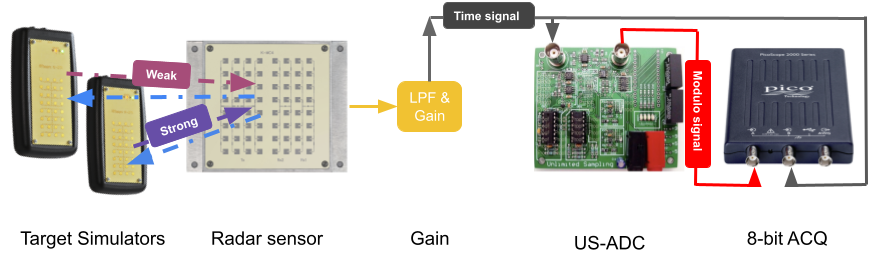}
\vspace{-0.15cm}
    \caption{\footnotesize{Experimental test-bed for the US-radar.}}
    \label{fig:lab}
    \vspace{-0.2cm}
\end{figure}

That said, the true benefits and potential of an end-to-end USF-enabled radar technology remain to be established.

\bpara{Contributions.} The main contribution of this paper is to present the core benefits of employing USF for radar signal processing. 
Our contributions include, 
\begin{enumerate}[leftmargin=*, label = $\mathrm{C}_\arabic*)$]
\item  The first end-to-end US-Radar system based on custom-designed hardware architecture.
\item Experiments based on our test-bed (see Fig.~\ref{fig:lab}) show that our recovery method is able to estimate the velocity of targets whose amplitudes are below the quantization noise floor of conventional ADCs. In particular, we observe that the noise floor has been lowered by $10$ dB (given the same bit-budget) in comparison to a conventional radar.
\end{enumerate}
\vspace{-0.2cm}
\section{Notations}\label{sec:not}
\vspace{-0.25cm}
Vectors are denoted in bold (\eg $\bs r$), the finite difference of order $N$ is defined as $\Delta^N a [k] = \Delta ( \Delta^{N-1} a [k]) $, with $\Delta a[k] = a[k+1] - a[k]$.
A random variable following $\mathcal{U}(a,b)$ follows a uniform distribution between $a$ and $b$.
$\mathcal{H}_K(\cdot)$ is the hard-thresholding operator that nulls all but the $K$ biggest components in amplitude. The operator $\S$ is the antidifference operator that defines the inverse of $\Delta$.    
\vspace{-0.2cm}
\section{System Model}\label{sec:model}
\vspace{-0.25cm}
In this paper, we consider a mono-frequency Doppler radar with one transmit and one receive antenna.

Considering $K$ targets located at ranges $R_k$ and with respective radial velocities $v_k$, the demodulated Doppler signal can be expressed as
\begin{equation}\label{eq:radarmodel}\textstyle
    r(t)= \sum\limits_{k=1}^K \alpha_k \cos(2 \pi f_D^k + \theta^k),
\end{equation}
with $\alpha_k$ being the amplitude, $f_D^k$ the Doppler frequency $f_D^k = 2 v_k f_0/\mathsf{c}$, and $\theta^k$ captures the phase resulting from both the range $R$ and the demodulation process.
$f_0$ is the operating frequency of the radar, and $\mathsf{c}$ is the speed of light.
Section \ref{sec:model} introduced the radar signal model; this signal, however, still needs to be digitized using an \textit{Analog to Digital Converter} (ADC) before any processing can be done.

In this paper, we consider the classic ADC structure that is the mid-rise quantizer with $b$ bits and a dynamic range of $\pm \beta$, which can be expressed as $\mathcal{Q}_b^{\beta}(a) =(\lfloor \frac{a}{\delta_\beta} \rfloor + \frac{1}{2}) \delta_\beta$,
with the range bin resolution of the ADC defined as $\delta_\beta = \frac{2\beta}{2^b -1}$.
\begin{figure*}
    \centering
    \includegraphics[width=0.75\textwidth]{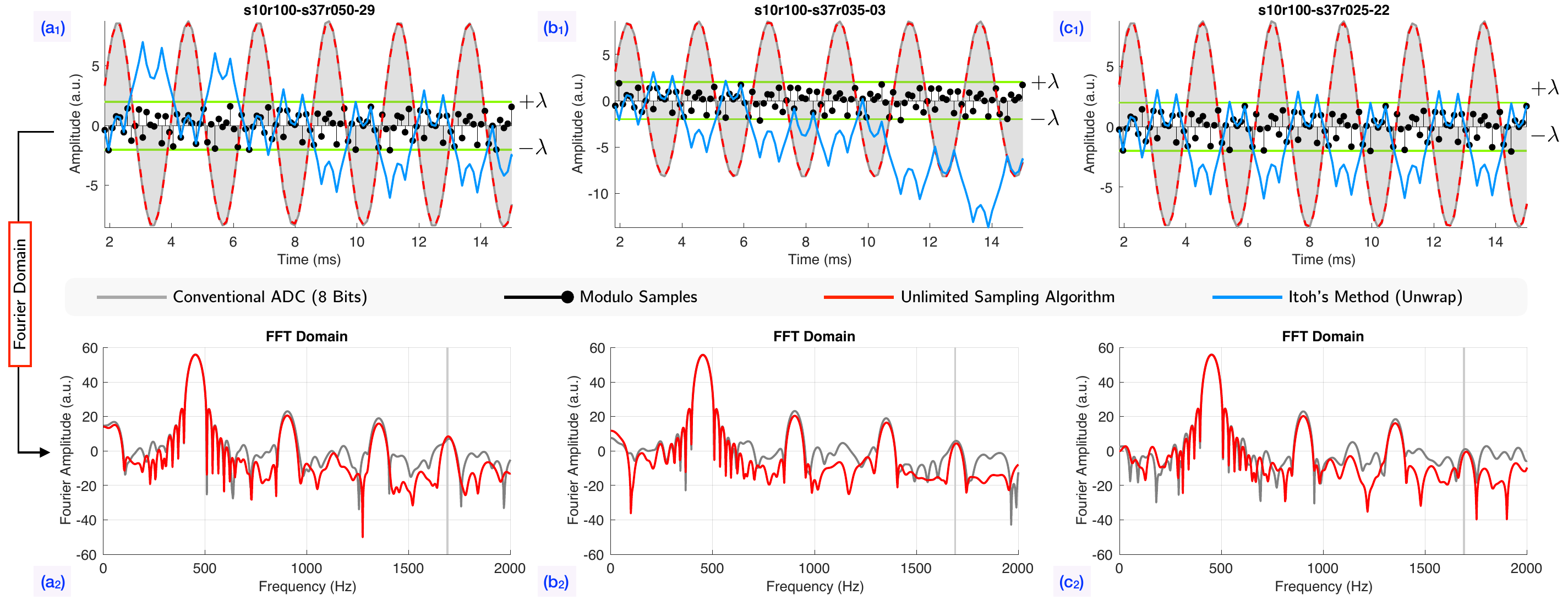}
\caption{\footnotesize{Experimental results using the test-bed represented in Fig.\ref{fig:lab}, The $3$ columns represent $3$ different experiment setups where the amplitude of the \textit{weak} target is varied.
The first row represents the time signal acquired at a sampling rate of $8140$Hz and their corresponding spectrum on the second row.
Each colour of curve represents a different acquisition and reconstruction strategy, namely grey for the standard mid-rise quantizer, black for the modulo samples, blue for the reconstruction using \textit{unwrap}, and red for the one provided by \cite{Bhandari:2019:Ja}.}}
    \label{fig:exp}
\end{figure*}
The quantizer defined 
induces, because of the finite nature of the digitization process itself, imperfections in the acquired signals. 
Indeed, saturation occurs when the signal's dynamic range exceeds the one of the ADC ($|a|>\beta$) and quantization noise arises from the $\delta_r$-discretization of the signal.
For a fixed number of bits $b$, there exist no single solutions to mitigate both of these effects simultaneously.

In this paper, we focus on a particular radar scenario where, because of the radar equation \cite{skolnik1980} the power of targets varies as $\mathcal{O}(\sigma R^{-4})$, with $R$ the range of the target and $\sigma$ its \textit{Radar Cross Section} (RCS).
Furthermore, the RCS of targets, \ie the power that they reflect, varies dramatically with the material, geometry and orientation at which the target is observed \cite{skolnik1980}. 
In the context of this \textit{Near-Far} recovery problem with a \textit{strong} and a \textit{weak} target, balancing between saturation and quantization noise is of paramount importance to detect both targets.
\vspace{-0.4cm}
\section{Unlimited Sampling}\label{sec:US}
\vspace{-0.25cm}
In this paper, we propose to do away with the limitations of the previously introduced acquisition strategy by leveraging a co-design approach between the acquisition and reconstruction known as Unlimited Sampling.
Unlimited Sampling, which was introduced in \cite{Bhandari:2017:C}, proposes to acquire a folded version of the signal of interest and to leverage its band-limited structure to reconstruct it. 
The folding operator is defined as
 $   \MO{r} = r+\lambda - \left\lfloor \frac{r+\lambda}{2\lambda}
\right\rfloor  2 \lambda$,
in words, the operator $\MO{\cdot}$ folds the signal around $\pm \lambda$. 
Our measurement model (considering a sampling rate $\frac{1}{T}$) can thus be expressed as
  $  y[n] = \mathcal{Q}_b^{\lambda}(\MO{r(n T)})$. 

An iterative algorithm and the theoretical setting in which perfect reconstruction is achieved are provided in \cite{Bhandari:2020:Ja}. 
This shows that the necessary sampling rate for US-based techniques is simply a multiple, by $2 \pi e$, of the one given by Shannon's sampling theorem. 
\vspace{-0.25cm}
\section{Lab Experiment} \label{sec:lab}
\vspace{-0.2cm}
Our experimental setup is the following: 
a commercial radar sensor emits a continuous wave that is then reflected by two Doppler simulators; these two echoes are then captured by the radar's receiving antenna and are acquired, after a filtering stage, by the USF-enabled ADCs (see Fig.\ref{fig:lab}).
The radar sensor \cite{KMC4} consists of one transmitting antenna and two receiving antennas connected to a coherent demodulator, operated as a Doppler radar with a centre frequency of $24$GHz.
By capturing, modulating and re-emitting the signal transmitted by the radar sensors, KDT-1 simulators (from RF-Beam \cite{KDT1}) are able to simulate moving targets in the K-band up to $\pm 200$km/h.
We use the modulo ADC \cite{Bhandari:2021:J} with $\lambda = 2.01$V. 
To fully leverage the potential of the \textit{unlimited} dynamic range offered by the USF, the radar signal is amplified, using a filtering stage, to cover a range of $\pm 8$V and the dynamic range of the $8$-bit ADC \cite{pico} is set to fit the modulo acquisition.
To compare the results of this acquisition, a second ADC is used with the dynamic range set to $\beta = 8$V.

The results of our experimental setup are compiled in Fig.\ref{fig:exp}.
The spectrum is made of a peak in $450$Hz, corresponding to the \textit{strong} target at $10$km/h, and a second \textit{weaker} one at $1650$Hz ($37$km/h). 
The target simulators suffer from slight imperfections that generate harmonics; these can be observed for the higher power target at $900$ and $1350$Hz.
In this first measurement, in Fig.\ref{fig:exp}($a_1$, $a_2$), the peak of the \textit{weak} target is easily distinguishable from the noise floor for the classic acquisition (in gray).
In Fig.\ref{fig:exp}$a_1$), one can compare the reconstruction provided by US algorithm
(in red) and the classic \textit{unwrap} method. 
The latter is not able to successful reconstruction. 
Indeed, methods like \textit{unwrap} are limited by the high sampling rate they require, compared to USF, as they rely on Itoh's condition \cite{Itoh1982AnalysisOT}.
Lowering the power allocated to the \textit{weak} target (see Fig.\ref{fig:exp}$b_2$-$c_2$), one can observe, for the classic acquisition, that the peak corresponding to the \textit{weak} is starting to fall below the quantization noise which hinders its detection. 
The higher sensitivity of the Unlimited Sampling methods is clearly highlighted here as the peaks in Fig.\ref{fig:exp}($b_2$-$c_2$) are still distinguishable from the lowered quantization noise. 
In this experimental setup, the resolution gain between $\beta = 8$V and $\lambda = 2.01$ corresponds to a gain of $2$ bits, while the same $8$-bit ADC was used. 
As a consequence, the noise floor has been lowered by $10$dB.
Herein lies the power of Unlimited Sampling when applied in practical settings; removing the constraint on the dynamic range, thanks to the modulo ADC, enables higher quality reconstruction without increasing the bit-budget devoted to the acquisition.   
\vspace{-0.35cm}
\section{Conclusion}\label{sec:conclusion}
\vspace{-0.2cm}
In the paper, an end-to-end study of the practical use of Unlimited Sampling for increasing the sensitivity of a radar sensor was presented. 
Leveraging the co-design approach between hardware and processing, we showed
through measurements, that acquiring signals from different targets with widely varying amplitude is challenging for classic ADCs with fixed resolution. 
Using the Unlimited Sampling framework we showed that we are able to lower the noise floor coming from the quantization noise while simultaneously avoiding saturating the measured signal. 
The measurement setup using both a commercial radar sensor and a USF-enabled ADC showed that the estimation of targets below the noise floor of classic ADCs is made possible without increasing the resolution of the acquisition.
The reconstructed spectrums using this strategy exhibit a noise floor lowered by more than $10$dB.

\bibliographystyle{IEEEtran}
\bibliography{IEEEabrv,bib_TF,ABList_2022}

\end{document}

%% file: preamble_2.tex
\usepackage{tikz}
\usepackage{url}
\usepackage{cancel}
\usepackage{blindtext}
\usepackage{graphicx}
\usepackage{pgfplots}
\usepackage{hyperref}
\usepackage{color,colortbl}
\usepackage{xspace}
\usepackage{epstopdf}
\usepackage{amssymb}
\usepackage[cmex10]{amsmath}
\usepackage{xcolor}
\usepackage[normalem]{ulem}
\usepackage[footnotesize]{caption}
\usepackage{subcaption}
\usepackage{amsthm}
\usepackage{cite}

\usepackage[utf8]{inputenc}
\usepackage{pgfplots}
\DeclareUnicodeCharacter{2212}{−}
\usepgfplotslibrary{groupplots,dateplot}
\usetikzlibrary{patterns,shapes.arrows}
\pgfplotsset{compat=newest}

\newcommand{\bs}{\boldsymbol}

\newcommand{\ie}{\emph{i.e.}, \xspace }
\newcommand{\eg}{\emph{e.g.},\xspace }

\newcommand{\iid}{%
    \ifmmode
        \mathrm{i.i.d.}%
    \else%
        i.i.d.\@\xspace%
    \fi%
}

\usetikzlibrary{through,calc}

\definecolor{bblue}{HTML}{4F81BD}
\definecolor{rred}{HTML}{C0504D}
\definecolor{ggreen}{HTML}{9BBB59}
\definecolor{ppurple}{HTML}{9F4C7C}

\definecolor{Liberty}{HTML}{4357AD}
\definecolor{Burlywood}{HTML}{48A9A6}
\definecolor{FuzzyWuzzy}{HTML}{C1666B}

\definecolor{Worange}{HTML}{D4B483}

\definecolor{mDarkBrown}{HTML}{C1666B}
\colorlet{mMediumBrown}{mDarkBrown!80}
\colorlet{mLightBrown}{mDarkBrown!50}

\definecolor{mDarkTeal}{HTML}{4357AD} 
\colorlet{mMediumTeal}{mDarkTeal!80}
\colorlet{mLightTeal}{mDarkTeal!50}

\colorlet{redn}{mDarkBrown}
\colorlet{bluen}{mDarkTeal}

\usepackage[colorinlistoftodos]{todonotes}

\definecolor{TF}{HTML}{008080}

\usepackage[english]{babel}

%% file: AB_Preamble.tex
\usepackage{nicefrac,mathrsfs,bbm}

\newcommand{\bpara}[1]		{\smallskip \noindent {\bf #1}}
\usepackage[inline, shortlabels]{enumitem}
\setlist{itemjoin ={,\enspace},itemjoin* = {, and\enspace}}

\newcommand{\ft}[1]			{\left[\kern-0.15em\left[#1\right]\kern-0.15em\right]}
\newcommand{\fe}[1]		{\left[\kern-0.30em\left[#1\right]\kern-0.30em\right]}

\newcommand{\MO}[1]		{\mathscr{M}_\lambda ({#1} )}

\renewcommand\bar\underline
\renewcommand\hat\widehat
\renewcommand\geq\geqslant
\renewcommand\leq\leqslant

\def\S					{\mathsf{S}}